\documentclass[fleqn,usenatbib]{aa}

\usepackage{graphicx}
\usepackage{ifthen}
\usepackage{url}
\usepackage{amsmath}
\usepackage{bm}
\usepackage{lscape}
\usepackage{rotating}
\usepackage[graphicx]{realboxes}
\usepackage{supertabular}
\usepackage{pdfpages}
\usepackage{footnote}
\usepackage{hyperref}
\usepackage{amssymb}
\makesavenoteenv{tabular}
\makesavenoteenv{table}

\usepackage[T1]{fontenc}
\usepackage{tipa}
\usepackage{ulem}

\hypersetup{
  colorlinks   = true,    
  urlcolor     = blue,    
  linkcolor    = blue,    
  citecolor    = blue      
}

\def\kms{{\rm\,km \  s^{-1}}}

\def\kpc{{\rm\,kpc}}

\def\msun{{\rm\,M_\odot}}

\def\teff{{\rm\,T_{eff}}}
\def\vmicro{{\rm\,v_{micro}}}

\def\eg{{e.g.{} }}

\def\ione{{~\sc i}}
\def\ii{{~\sc ii}}

\usepackage{amsmath}

\def\FeH{{\rm[Fe/H]}}

\begin{document}

\title{The Pristine Inner Galaxy Survey (PIGS) IX}
\subtitle{The largest detailed chemical analysis of very metal-poor stars in the Sagittarius dwarf galaxy}

\author{Federico Sestito\inst{1}
\and Sara Vitali\inst{2,3}
\and Paula Jofre\inst{2,3}
\and Kim A. Venn\inst{1}
\and David S. Aguado\inst{4,5}
\and Claudia Aguilera-G\'omez\inst{6}
\and Anke Ardern-Arentsen\inst{7}
\and Danielle de Brito Silva\inst{2,3}
\and Raymond Carlberg\inst{8}
\and Camilla J. L. Eldridge\inst{2,3}
\and Felipe Gran\inst{9}
\and Vanessa Hill\inst{9}
\and Pascale Jablonka\inst{10,11}
\and Georges Kordopatis\inst{9}
\and Nicolas F. Martin\inst{12,13}
\and Tadafumi Matsuno\inst{14}
\and Samuel Rusterucci\inst{12,15}
\and Else Starkenburg\inst{15}
\and Akshara Viswanathan\inst{15}
}

\institute{Department of Physics and Astronomy, University of Victoria, PO Box 3055, STN CSC, Victoria BC V8W 3P6, Canada\\
\email{sestitof@uvic.ca}
\and
Instituto de Estudios Astrof\'isicos, Universidad Diego Portales, Av. Ej\'ercito Libertador 441, Santiago, Chile\\
\email{sara.vitali@mail.udp.cl}
\and
Millenium Nucleus ERIS
\and
Instituto de Astrof\'isica de Canarias, E-38205 La Laguna, Tenerife, Spain
\and
Universidad de La Laguna, Departamento de Astrof\'isica, 38206 La Laguna, Tenerife, Spain
\and
Instituto de Astrof\'isica, Pontificia Universidad Cat\'olica de Chile, Av. Vicuna Mackenna 4860, Santiago, Chile
\and
Institute of Astronomy, University of Cambridge, Madingley Road, Cambridge CB3 0HA, UK
\and
Department of Astronomy \& Astrophysics, University of Toronto, Toronto, ON M5S 3H4, Canada
\and
Universit\'e Cote d'Azur, Observatoire de la Cote d'Azur, CNRS, Laboratoire Lagrange, Nice, France
\and
Laboratoire d'astrophysique, \'Ecole Polytechnique F\'ed\'erale de Lausanne (EPFL), Observatoire, CH-1290 Versoix, Switzerland
\and
GEPI, Observatoire de Paris, Universit\'e PSL, CNRS, 5 Place Jules Janssen, F-92195 Meudon, France
\and
Universit\'e de Strasbourg, CNRS, Observatoire astronomique de Strasbourg, UMR 7550, F-67000 Strasbourg, France
\and
Max-Planck-Institut fur Astronomie, Königstuhl 17, D-69117 Heidelberg, Germany
\and
Astronomisches Rechen-Institut, Zentrum für Astronomie der Universität Heidelberg, Mönchhofstr. 12-14, 69120 Heidelberg, Germany
\and
Kapteyn Astronomical Institute, University of Groningen, Landleven 12, 9747 AD Groningen, The Netherlands
}

\date{Accepted XXX. Received YYY; in original form ZZZ}

\abstract{
The most metal-poor stars provide valuable insights into the early chemical enrichment history of a system, carrying the chemical imprints of the first generations of supernovae.
The  most metal-poor region of the Sagittarius dwarf galaxy remains inadequately observed and characterised. To date, only $\sim4$  stars with $\FeH<-2.0$ have been chemically analysed with high-resolution spectroscopy. In this study, we present the most extensive chemical abundance analysis of 12 low-metallicity stars with metallicities down to $\FeH=-3.26$ and located in the main body of Sagittarius.
These targets, selected from the Pristine Inner Galaxy Survey, were observed using the MIKE high-resolution spectrograph at the {\it Magellan-Clay} telescope, which allowed us to measure up to 17 chemical species. The chemical composition of these stars reflects the imprint of a variety of type~II supernovae (SNe~II). A combination of low- to intermediate-mass high-energy SNe and hypernovae ($\sim10-70\msun$) is required to account for the abundance patterns of the lighter elements up to the Fe-peak. The trend of the heavy elements suggests the involvement of compact binary merger events and fast-rotating (up to $\sim300\kms$) intermediate-mass to massive metal-poor stars ($\sim25-120\msun$) that are the sources of rapid and slow processes, respectively. Additionally, asymptotic giant branch stars contribute to a wide dispersion of [Ba/Mg] and [Ba/Eu]. The absence of an $\alpha-$knee in our data indicates that type Ia supernovae did not contribute in the very metal-poor region ($\FeH\leq-2.0$). However, they might have started to pollute the interstellar medium  at $\FeH>-2.0$, given the relatively low [Co/Fe] in this metallicity region.}


\maketitle

\section{Introduction}
The formation and evolution of galaxies is significantly influenced by the merging of systems and the ingestion of smaller galaxies into larger ones \citep{White78,Frenk88,NavarroFrenkWhite97}. The Sagittarius (Sgr) dwarf galaxy \citep{Ibata1994} serves as an ideal laboratory for testing models of chemical evolution and hierarchical accretion dynamics. The system is estimated to have experienced its first in-fall about $5$ Gyr ago \citep[\eg][]{RuizLara20}, and it is located approximately 26.5 kpc  away from us towards the inner Galactic regions  \citep{Vasiliev20}. Sgr represents the third most massive satellite galaxy in the Local Group (total mass of $\sim4.8\cdot 10^8 \msun$), following only the two Magellanic Clouds \citep[\eg][]{Monaco04,Mcconnachie12,Vasiliev20}. 

As it is being tidally stripped by the Milky Way (MW), the system is now visible with its core and two stellar streams \citep{Ibata1994,Mateo98,Majewski03,Law10,Belokurov14}, dragging with it several globular clusters close to its core (M54) and to the stellar streams \citep[Terzan 7, Terzan 8, Arp 2, and Palomar 12,][]{Sbordone07,Mucciarelli17}. Several studies have investigated its prolonged and complex star formation history (SFH), characterised by multiple star formation episodes, employing various techniques such as high-resolution spectroscopy \citep[\eg][]{Bonifacio00,Monaco05,Chou07,McWilliam13,Hansen18Sgr,Hayes20,Hasselquist17,Hasselquist21} and photometric  techniques \citep[\eg][]{Bellazzini99,Layden00,Siegel07,Vitali22},  revealing the presence of at least four distinct stellar populations \citep{Siegel07}. According to the work of \cite{Siegel07} these populations are divided into the oldest and most metal-poor stars (with $\FeH\lesssim-1.0$ and ages exceeding 10 Gyr), the intermediate group with ages between 4 to 8 Gyr and $\FeH\sim-0.6$,  a young population  of $\sim2.5$ Gyr old and $\FeH\sim-0.1$, and the youngest stars exhibiting super-solar metallicities ($\FeH\sim+0.5$, $<2$ Gyr).

Sgr's core is dominated by metal-rich stars  due to the recent star formation \citep{Siegel07} and to the fact that the oldest and metal-poor population has been preferentially stripped by tidal perturbations \citep{Monaco05,Monaco07,Carlin18,Ramos22}. Additionally, Sgr stars overlap  in the colour-magnitude diagram with the Galactic bulge \citep{Monaco05,Mucciarelli17}, hence  exploring its most metal-poor tail has posed significant challenges so far. To date, only approximately four very metal-poor stars (VMPs, $\FeH\leq-2.0$) have been analysed with high-resolution spectroscopy, providing chemical abundances for various elements \citep{Hansen18Sgr}, while \citet{Chiti19} and \citet{Chiti20Sgr} measured only metallicities and carbonicities ([C/Fe]) for 11 VMPs. On the other hand,  the chemo-dynamical properties of the metal-poor population of Sgr has only been studied in its stellar streams, with a large effort by  APOGEE  \citep{Hayes20} and by the H3 survey \citep{Johnson20}.  Consequently, three decades after its discovery \citep{Ibata1994}, the VMP tail of Sgr's core remains largely unexplored. Yet, this metallicity region is crucial to gain insights into the early chemical enrichment of Sgr, as these stars represent relics of the system's initial stellar population.

The synergy between the exquisite astrometry and photometry from {\it Gaia} \citep{Gaia16,GaiaEDR3,GaiaDR3} and metal-poor dedicated photometric surveys such as Pristine \citep{Starkenburg17b,Martin23} offers the most efficient way of probing the most metal-poor tail of Sgr. While Pristine primarily focusses on the Galactic halo \citep[see spectroscopic works from][]{Starkenburg18,Bonifacio19,Aguado19,Venn20,Kielty21,Lucchesi22,Martin22,Yuan22}, its sub-survey, the Pristine Inner Galaxy Survey (PIGS),  targets  the VMP candidates towards the Galactic bulge \citep{Arentsen20a,Arentsen24,Mashonkina23,Sestito23,Sestito24gh} and Sgr \citep{Vitali22}. Both surveys use a narrow-band $\rm Ca\,H\&K$ filter, sensitive to metallicity, to derive photometric metallicities. The subsequent medium-to-high-resolution spectroscopic follow-up of PIGS VMP candidates proved the powerful efficiency of the Ca~HK filter even in high-extinction regions \citep{Arentsen20b}. 

Recently, \citet{Vitali22}, thanks to the metallicity-sensitive photometry obtained from the Pristine Ca~HK filter, investigated the metallicity distribution of $\sim50,000$ Sgr candidate members as a function of their spatial position. 
This study, encompassing the largest sample of VMP candidate members in Sgr ($\sim1200$ stars), revealed a negative photometric metallicity gradient extending up to $\sim5.5\kpc$  from Sgr's centre. This metallicity gradient has been recently quantified to be $-2.48\lesssim \nabla\rm{[M/H]} (10^{-2}) \lesssim-2.02 $~dex~deg$^{-1}$ \citep{Cunningham24}. The metallicity gradient suggests outside-in star formation, indicating that metal-poor stars formed throughout the system, whereas more metal-rich stars are concentrated in the inner regions. Similar observational features have been observed in other classical dwarf galaxies \citep[\eg][]{Tolstoy04,Battaglia08b,Zhang12,Sestito23scl,Sestito23Umi,Tolstoy23} and modelled with cosmological simulations \citep[\eg][]{Revaz18}. 

This work constitutes the most extensive chemical analysis of VMPs in Sgr to date. Stars have been selected within PIGS from the study of \citet{Vitali22} and observed with the MIKE high-resolution spectrograph at the {\it Magellan-Clay} Telescope \citep{2003Bernstein}. Target selection is outlined in Section~\ref{sec:data}, while Section~\ref{sec:modelatm} describes the model atmosphere analysis. A comparison with the metallicities and radial velocity from the PIGS medium-resolution analysis is outlined in Section~\ref{sec:compAAT}. The results and discussion on the chemical properties of the VMP tail of Sgr are reported in Section~\ref{sec:discussion}. Conclusions are summarised in Section~\ref{sec:conclusion}.

\section{Data}\label{sec:data}
\subsection{Target selection from the Pristine Inner Galaxy Survey}

A low- and medium-resolution spectroscopic follow-up was conducted for 13,000 PIGS very metal-poor candidates \citep{Arentsen20b}, which were observed with the AAOmega+2dF mounted on the Anglo Australian Telescope (AAT, \citealt{Saunders2004,2002Lewis,Sharp2006}). Four AAT pointings were allocated to observe stars in the Sagittarius dwarf galaxy \citep{Vitali22, Arentsen24}. From their observed spectra, metallicities, radial velocities, carbonicities, and stellar parameters were derived using spectrum fitting techniques, employing  \textsc{FERRE}\footnote{\url{http://github.com/callendeprieto/ferre}} \citep{Prieto2006,Aguado2017b}. 

The targets for this study were chosen from the PIGS/AAT spectroscopic catalogue, ensuring their Sgr membership based on their {\it Gaia} proper motion and photometry and from their AAT radial velocities  as outlined in \citet{Vitali22}. Briefly, this selection consists in {\it Gaia} stars with reduced proper motion within a radius of $0.6$~mas~yr$^{-1}$ as in \citet{Vasiliev20} and radial velocities in the range $100-200\kms$ \citep[\eg][]{Ibata1994,Bellazzini2008}. Additionally, VMP stars with $\FeH_{\rm AAT} \leq-2.0$ were selected from the Sgr core, excluding members of the nearby  nuclear star cluster M54. The limiting magnitude was set to $G = 16$ mag. Subsequently, stars for which their AAT spectrum has S/N~$> 50$ in the calcium triplet region were selected.

\begin{figure}[h]
\centering
 \includegraphics[width=\columnwidth]{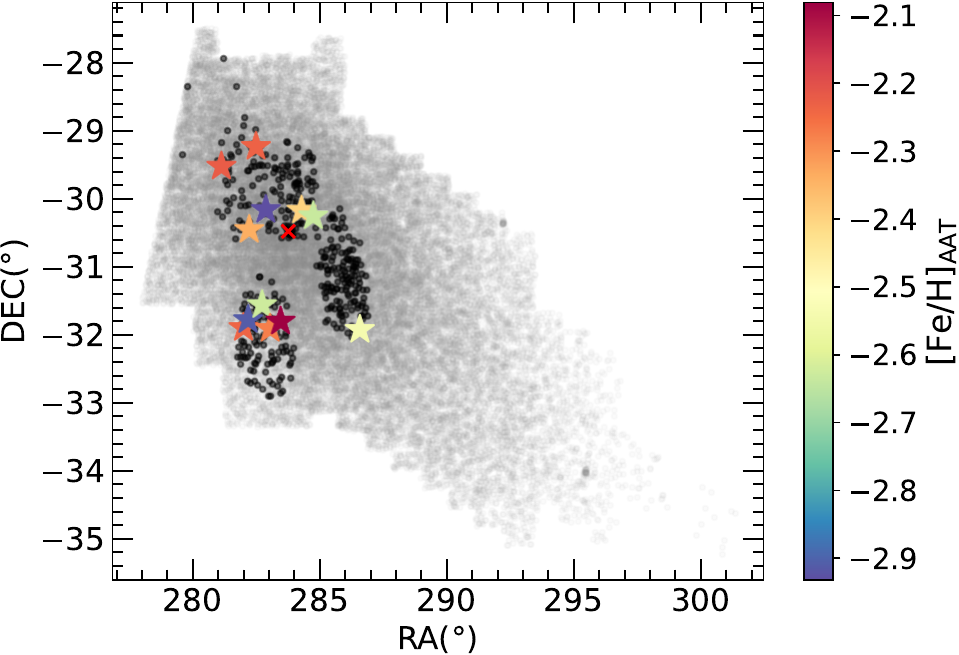} 
 \caption{Sgr core in equatorial coordinates. The 12 targets are denoted by a star marker, colour-coded by their  metallicity from PIGS/AAT. Candidate Sgr members in the PIGS photometric footprint, following \citet{Vitali22}, are marked by grey dots. Black circles denote Sgr stars observed with AAT low-/medium-resolution \citep{Sestito24Sgr2}. The red cross marks the position of the nuclear globular cluster M54 (NGC6715).}
 \label{fig:targets}
\end{figure}

Candidate members of Sgr in the PIGS photometric footprint (grey dots) identified as in \citet{Vitali22}, the Sgr members with AAT spectra (black circles), and the 12 selected targets colour-coded by their PIGS/AAT metallicities are displayed in Figure~\ref{fig:targets}.  Table~\ref{tab:targets} reports the {\it Gaia} DR3 source ID, coordinates, magnitudes, and the reddening of our targets.

\subsection{Observations and spectra reduction}
The observations were conducted over two runs using Chilean time spanning from 2022 to 2023 (CN2022A-30, CN2023A-62 P.I.: S. Vitali), utilising the MIKE echelle spectrograph \citep{2003Bernstein} installed on the {\it Magellan-Clay} telescope. The use of the 0.7" slit and 2x2 binning ensured a resolving power of $R\sim 35,000$ in the blue and $R \sim 28,000$ in the red. While the wavelength range of the MIKE spectra covers the range $\sim 3500-9000$ \AA,  the signal-to-noise ratio (S/N) restricts the usable  spectra to wavelengths $>3850$~\AA. 

Data reduction was performed utilising the default \texttt{CarPy} pipeline \citep{2000Kelson,2003Kelson}, which was tailored for MIKE spectra\footnote{\url{https://code.obs.carnegiescience.edu/mike}}. Following reduction, individual spectral orders were extracted, merged, flux normalised, and subsequently corrected for barycentric and radial velocities. Radial velocities are measured cross-correlating the observed normalised spectra with metal-poor templates generated from \textsc{OSMARCS} stellar atmosphere models \citep{Gustafsson08,Plez12}.

\begin{figure*}[h!]
\includegraphics[width=\textwidth]{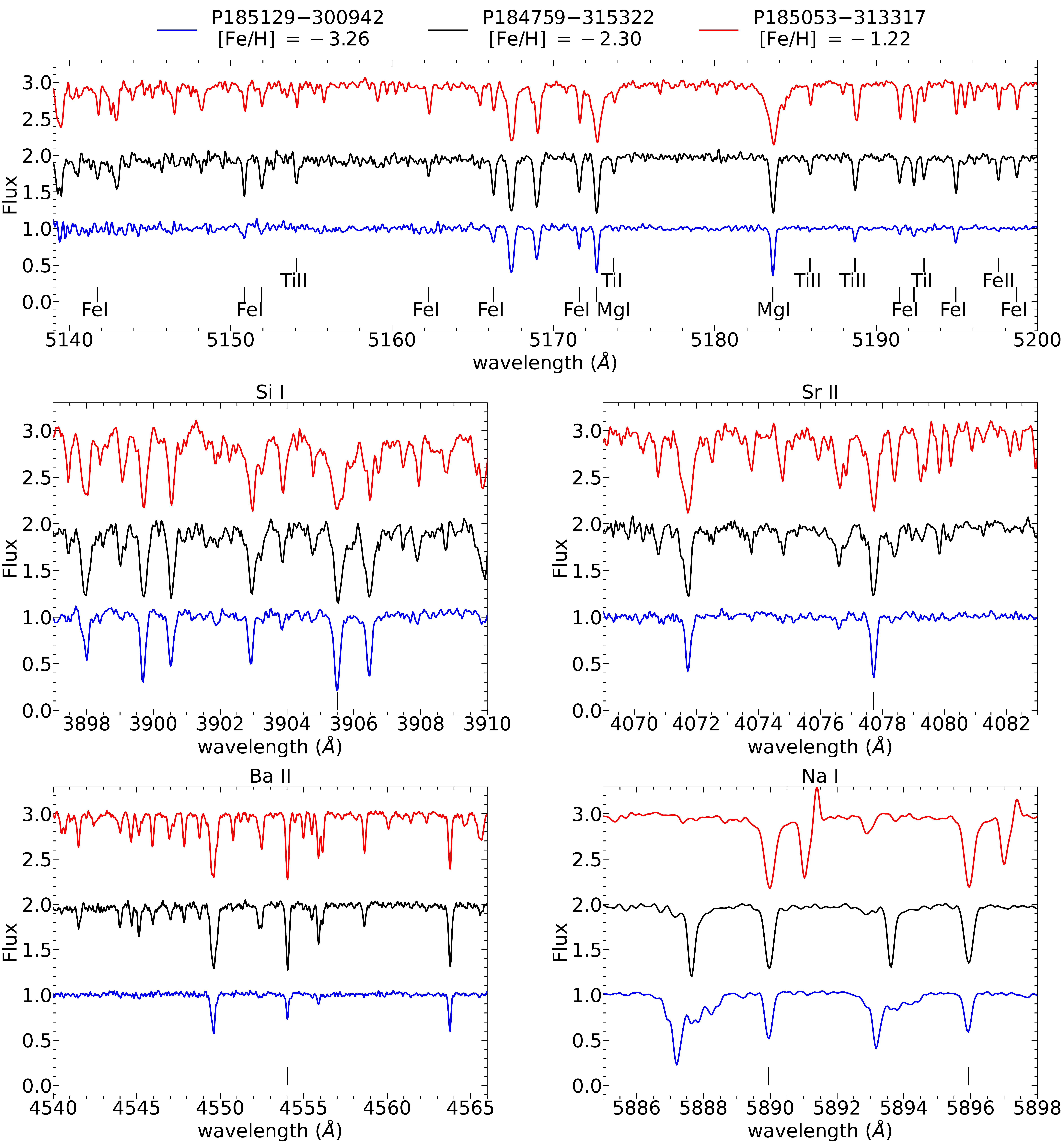}
\caption{Example of MIKE spectra. Top panel: Mg\ione{} Triplet region, which also includes several Fe and Ti lines. Vertical  short lines marks the position of Mg\ione{}, Ti, and Fe lines found in most of the targets. Central left panel: Si\ione{} 3905.523 \AA{} line region. Central right: Sr\ii{} 4077.709 \AA{} line region.  Bottom left: Ba\ii{} 4554.029 \AA{} line region. Bottom right: Na\ione{} Doublet $\lambda\lambda 5889.951, 5895.924$ \AA{} lines region, which also includes spectral lines from various clouds of the interstellar medium. The S/N of these spectra, measured in the Mg\ione{}b Triplet region, ranges from $\sim24$ to $\sim56$. Vertical short lines in the central and bottom panels indicate the position of the Si (central left), Sr (central right), Ba (bottom left), and Na (bottom right) spectral lines. }
\label{Fig:spectra}
\end{figure*}

Examples of reduced and normalised spectra are shown in Figure~\ref{Fig:spectra}. Each panel displays the MIKE spectra of the most metal-poor star (P185129$-$300942), the most metal-rich (P185053$-$313317), and a VMP target (P184759$-$315322) from our sample. The panels in Figure~\ref{Fig:spectra} show the Mg~I~ triplet region (top panel, $5140-5200$ \AA), the Si\ione{} $3903$ \AA{} spectral line (centre left), the Sr\ii{} $4078$ \AA{} line (centre right), the Ba\ii{} $4554$ \AA{} (bottom left), and the Na\ione{} doublet $5890,5896$ \AA{} lines (bottom right).

Table~\ref{tab:targets} also provides details on the exposure time, the number of observations, and the S/N for each target. A note on the exposure time, it is important to consider that Sgr's core is located in a region with high extinction, which significantly affects the bluer part of the stellar spectrum. Therefore, the relatively long exposure is needed to achieve the desired S/N even in the bluest spectral regions. The obtained S/N are comparable to other works targeting high-extinction regions, such as the MW bulge \citep[\eg][]{Reggiani20}, with MIKE. Radial velocities are reported in Table~\ref{tab:stellar}.

\begin{table*}
\caption[]{Pristine names, {\it Gaia} informations, and log of the observations.}
\centering
\resizebox{1\textwidth}{!}{
\begin{tabular}{lccccccccc}
\hline
Short name  & PIGS name &{\it Gaia} DR3 source id & $\alpha$ & $\delta$  &  G & BP$-$RP & A$_{\rm V}$ & T$_{\rm exp}$ & S/N \\
 &  & & (deg)& (deg)  &  (mag) & (mag)& (mag) & (s)  &  \\ \hline
P184431$-$293145 & Pristine\_184431.86$-$293145.0   & 4071285193071973504 & 281.13275 & -29.52916 &  15.18 & 1.54 & 0.36 & 6000 & 20 \\
P184759$-$315322 & Pristine\_184759.63$-$315322.5   & 6736966644012267648 & 281.99844 & -31.88958 & 15.41 & 1.45 & 0.37 & 6600 & 31\\
P184843$-$314626 & Pristine\_184843.24$-$314626.8   & 6736974929031902464 & 282.18017 & -31.77411 &  15.31 & 1.43 & 0.42 & 7200 & 51\\
P184853$-$302718 & Pristine\_184853.44$-$302718.4   & 6761132195931902208 & 282.22267 & -30.45511 &  15.00 & 1.60 & 0.52 & 5400 & 23\\
P184957$-$291425 & Pristine\_184957.04$-$291425.1   & 6761537434686446720 & 282.48767 & -29.24031 &  14.91 & 1.64 & 0.49 & 5400& 53\\
P185053$-$313317 & Pristine\_185053.71$-$313317.7   & 6736253885603451776 & 282.72379 & -31.55492 &  15.12 & 1.48 & 0.49 & 6000 & 24\\
P185129$-$300942 & Pristine\_185129.00$-$300942.8   & 6761245956700629504 & 282.87083 & -30.16189 &  15.79 & 1.37 & 0.45 & 9000 &  56\\
P185210$-$315413 & Pristine\_185210.30$-$315413.2   & 6736183692971769472 & 283.04292 & -31.90367 &  14.91 & 1.53 & 0.35 & 5400 & 48\\
P185347$-$314747 & Pristine\_185347.87$-$314747.6   & 6760197812879213056 & 283.44946 & -31.79656 &  14.89 & 1.61 & 0.42 & 5400 & 8\\
P185704$-$301021 & Pristine\_185704.51$-$301021.6   & 6760823641153762816 & 284.26879 & -30.17267 &  14.91 & 1.54 & 0.39 &5400 & 26\\
P185855$-$301522 & Pristine\_185855.01$-$301522.2   & 6760775163878990976 & 284.72921 & -30.25617 &  14.96 & 1.52 & 0.35 & 5400 & 26\\
P190612$-$315504 & Pristine\_190612.10$-$315504.4   & 6757188660021199360 & 286.55042 & -31.91789 &  15.21 & 1.42 & 0.28 & 6000 & 47\\
\hline
\end{tabular}
}
\tablefoot{PIGS name, the \textit{Gaia} DR3 source ID, the coordinates $(\alpha,\delta)$,  the \textit{Gaia} DR3 photometry G and BP$-$RP,  the reddening A$_{\rm V}$ from \citet{Schlegel98} and updated by \citet{Schlafly11}, the total exposure time, and the S/N at the Mg\ione{} Triplet region ($\sim5180$ \AA{}) reported for each target.}
\label{tab:targets}
\end{table*}

\section{Model atmosphere analysis}\label{sec:modelatm}

\subsection{Stellar parameters}\label{sec:stellparams}
The effective temperature and surface gravity were determined using the methodology outlined in details by \citet{Sestito23}. In summary, the  effective temperature was derived using a colour-temperature relation  \citep{Mucciarelli21}, akin to the Infrared Flux Method \citep[\eg][]{Gonzalez09},   adapted to the {\it Gaia} DR3 photometry. The surface gravity was computed using the Stefan-Boltzmann equation \citep[\eg][]{Kraft03,Venn17}, and an iterative process between effective temperature ($\teff$) and surface gravity (logg) was employed. The surface gravity depends on the heliocentric distance of these stars, which is assumed to be $26.5\pm2.5\kpc$ as they are located in Sgr \citep{Vasiliev20}. {\it Gaia} photometric magnitudes are de-reddened using the 2D dust map from \citet{Schlegel98} and updated by \citet{Schlafly11}. Then, the  {\it Gaia} extinction coefficients are derived using  $\rm A_V/E(B-V)= 3.1$ \citep{Schultz75} and the $\rm A_G/A_V = 0.85926$, $\rm A_{BP} /A_V = 1.06794$, $\rm A_{RP} /A_V = 0.65199$ relations \citep{Marigo08,Evans18}. Uncertainties associated with the stellar parameters were determined through Monte Carlo simulations applied to the input parameters. Although extinction towards Sgr has large values (see Table~\ref{tab:targets}), photometric effective temperatures provide a flat relation within the uncertainties between the Fe\ione{} lines abundances in Local Thermodynamic Equilibrium (LTE) and the excitation potential. Photometric effective temperatures are usually preferred over spectroscopic estimates as they are not affected by non-LTE effects \citep{Frebel13,Ezzeddine20}.

\begin{figure}[h]
\includegraphics[width=0.5\textwidth]{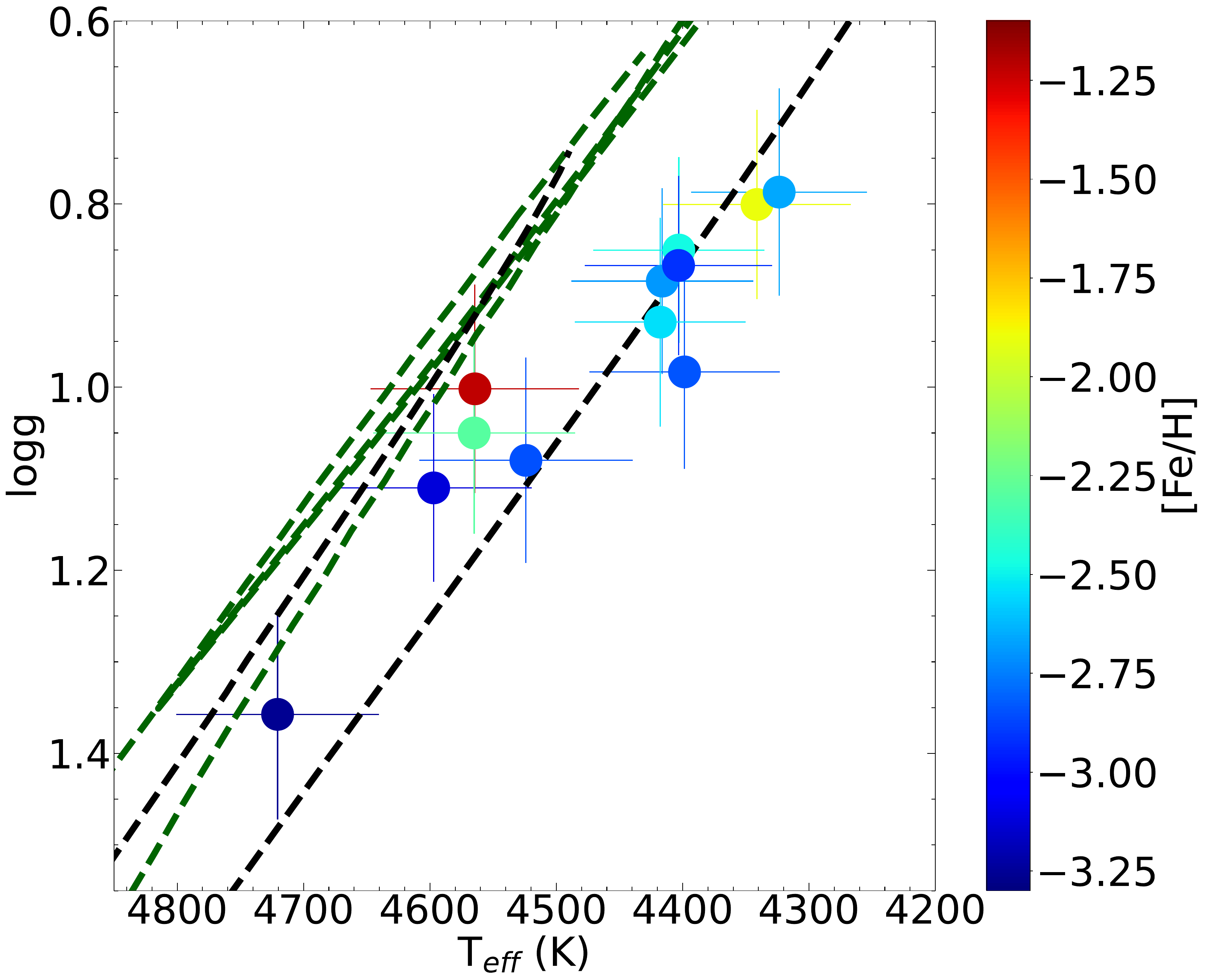}
\caption{Kiel diagram of the MIKE targets colour-coded by the MIKE metallicity of the targets. As a comparison,  an old very metal-poor isochrone from Padova  \citep[$\FeH = -2.0$, $11$ Gyr, dark green dashed line,][]{Bressan12} and two MESA/MIST isochrones \citep[$\FeH = -2.5, -2.0$, $12.6$ Gyr, black dashed lines,][]{Dotter16,Choi16}.}
\label{Fig:stellar}
\end{figure}

The Kiel diagram is displayed in Figure~\ref{Fig:stellar} with a very metal-poor isochrones from Padova \citep[dark olive dashed line,][]{Bressan12} and MESA/MIST \citep[black dashed lines][]{Dotter16,Choi16} as references. All the stars in this work are giants with effective temperature $4300< T_{\rm eff} <4700$ K and surface gravity $0.8<\rm{\log g} <1.4$. Table~\ref{tab:stellar} reports the effective temperature and surface gravity used in this work.

Initial estimates of the microturbulence velocity, $\vmicro$, were provided by \textsc{OSMARCS} models. These values were subsequently adjusted for some stars to ensure no dependency of the Fe\ione{} abundances as a function of the reduced equivalent width.

\begin{table}
\caption[]{Stellar parameters.}
\centering
\resizebox{0.5\textwidth}{!}{
\hspace{-0.6cm}
\begin{tabular}{lcccc}
\hline
PIGS name & RV  & T$_{\rm eff}$ & logg   & $\FeH$   \\
 & ($\kms$)  & (K) &   &     \\ \hline
P184431$-$293145 & 142.57 $\pm$ 0.67 & 4399 $\pm$ 75 & 0.98 $\pm$ 0.11 & $-2.84$ $\pm$ 0.05 \\
P184759$-$315322 & 184.00 $\pm$ 0.41 & 4565 $\pm$ 80 & 1.05 $\pm$ 0.11 & $-2.30$ $\pm$ 0.02 \\
P184843$-$314626 & 149.02 $\pm$ 0.71 & 4597 $\pm$ 77 & 1.11 $\pm$ 0.10 & $-3.12$ $\pm$ 0.03 \\
P184853$-$302718 & 121.41 $\pm$ 0.52 & 4403 $\pm$ 67 & 0.85 $\pm$ 0.10 & $-2.47$ $\pm$ 0.02 \\
P184957$-$291425 & 125.57 $\pm$ 0.74 & 4324 $\pm$ 69 & 0.79 $\pm$ 0.11 & $-2.66$ $\pm$ 0.02 \\
P185053$-$313317 & 153.21 $\pm$ 0.79 & 4564 $\pm$ 82 & 1.00 $\pm$ 0.11 & $-1.22$ $\pm$ 0.03 \\
P185129$-$300942 & 139.56 $\pm$ 0.77 & 4721 $\pm$ 80 & 1.36 $\pm$ 0.11 & $-3.26$ $\pm$ 0.02 \\
P185210$-$315413 & 170.25 $\pm$ 0.57 & 4416 $\pm$ 72 & 0.88 $\pm$ 0.10 & $-2.69$ $\pm$ 0.02 \\
P185347$-$314747 & 171.86 $\pm$ 0.69 & 4341 $\pm$ 74 & 0.80 $\pm$ 0.10 & $-1.91$ $\pm$ 0.02 \\
P185704$-$301021 & 140.96 $\pm$ 0.63 & 4403 $\pm$ 74& 0.87 $\pm$ 0.10 & $-2.91$ $\pm$ 0.06 \\
P185855$-$301522 & 134.67 $\pm$ 0.55 & 4418 $\pm$ 67 & 0.93 $\pm$ 0.11 & $-2.54$ $\pm$ 0.02 \\
P190612$-$315504 & 144.03 $\pm$ 0.81 & 4524 $\pm$ 84 & 1.08 $\pm$ 0.11 & $-2.85$ $\pm$ 0.02 \\
\hline
\end{tabular}}
\tablefoot{Radial velocities, effective temperatures, surface gravities, and metallicities  reported with their uncertainties.}
\label{tab:stellar}
\end{table}

\subsection{Spectral lines and atomic data}

The S/N of the spectra and the wavelength coverage of MIKE allow us to measure various species that are tracers of the early nucleosynthesis in Sgr. Carbon is measurable from the CH bands at $\sim4300$ \AA{} but only for 5 stars, since the spectrum is difficult to normalise due to the strength of both CH band itself and a Balmer line nearby.   Abundances of the $\alpha-$elements are derived from the spectral lines of Mg\ione, Si\ione, Ca\ione, Ti\ione{} and Ti\ii{}. Odd-Z element are present with Na\ione{}, Al\ione{}, K\ione{}, and Sc\ii{} lines. Fe-peak species show spectral lines of Fe\ione{} and Fe\ii, Cr\ione{}, Mn\ione{}, Co\ione{},  Ni\ione{}, and Zn\ione{}. The neutron-capture process elements detected in this work are Sr\ii{}, Y\ii{}, Ba\ii{}, and Eu\ii{}. The spectral line list used for these stars is available at the CDS.

Wavelengths of  spectral lines used in this work, their atomic data, hyper-fine structure (HFS) for Sc, Mn, and Co and isotopic corrections for Ba and Eu are sourced from \textsc{linemake} \citep{Placco21}. Molecular CH bands are from \citet{Masseron14} and corrections to [C/Fe] due to evolutionary effects are included as in \citet{Placco14}. Solar abundances are adopted from \citet{Asplund09}.

Chemical abundances are derived using \textsc{MOOG}\footnote{\url{https://www.as.utexas.edu/~chris/moog.html}}  \citep{Sneden73,Sobeck11}  from equivalent widths (EW) for most of the species. EW are  measured using \textsc{DAOSPEC} \citep{Stetson08}, which automatically fits  lines with Gaussian profiles. In case of very strong line depth, spectral lines have been fitted adopting a Voigt profile. Given the signal-to-noise ratio of our spectrum,  lines weaker than 20 m\AA{} are rejected.   \textsc{OSMARCS}\footnote{\url{https://marcs.astro.uu.se}} model atmospheres \citep{Gustafsson08,Plez12} have been employed, except for P184843$-$314626 and P185129$-$300942. The metallicities of these stars ($\FeH<-3.0$) are out of the \textsc{OSMARCS} ranges, so \textsc{Plez2000} model atmospheres were adopted instead. Synthetic spectra are fitted to observations using \textsc{MOOG} \textsc{synth} for the determination of [C/Fe] and [Eu/Fe]. The final chemical abundances are obtained from averaging the line by line abundances. Output files from \textsc{MOOG}, providing line-by-line chemical abundances, as well as a table summarising chemical ratios [X/H], [X/Fe], and their associated uncertainties are available at CDS.

\subsection{Uncertainties on the chemical abundances}
\textsc{MOOG} also provides the estimate of the chemical abundance scatter, $\delta_{\rm A(X)}$. The abundance uncertainties are calculated by combining the line-to-line scatter ($\delta_{\rm A(X)}$) with the uncertainties resulting from variations in the stellar parameters ($\delta_{\rm T_{eff}}$, $\delta_{\rm logg}$) in quadrature, leading to $\delta_{\rm A(X)}^{\rm tot}$. The final uncertainty for element X is given by $\sigma_{\rm A(X)}=\delta_{\rm A(X)}^{\rm tot}/\sqrt{{\rm N_X}}$, where N$_{X}$ is the number of lines measured for a given species. In case there is only one spectral line, the dispersion in A(Fe\ione{}) in the same spectral region is considered as the typical dispersion. This would give an idea of the internal dispersion due to the noise in that particular spectral region.

\subsection{Non-local Thermodynamic equilibrium corrections}\label{sec:nlte}
Non-local thermodynamic equilibrium corrections (NLTE) are applied to the [X/Fe] ratios only when these measurements are compared to supernovae yields models (see Section~\ref{sec:yields} and Figure~\ref{Fig:starfit}).

The MPIA database\footnote{\url{https://nlte.mpia.de}} is used to gather the NLTE corrections for Fe\ione{} and Fe\ii{} \citep{Bergemann2012}, Mg\ione{} \citep{Bergemann2017}, Si\ione{} \citep{Bergemann2013}, Ca\ione{} \citep{Mashonkina17}, Ti\ione{} and Ti\ii{} \citep{Bergemann2011}, Cr\ione{} \citep{Bergemann2010b}, Mn\ione{} \citep{Bergemann2019}, and Co\ione{} \citep{Bergemann2010}, while \textsc{INSPECT}\footnote{\url{http://www.inspect-stars.com}} is queried  for Na\ione{} \citep{Lind2012}. Ba\ii{} NLTE corrections are adopted from \citet{Mashonkina19}  online tool\footnote{\url{http://www.inasan.ru/~lima/pristine/ba2/}} and the corrections for K are from \citet{Ivanova00}.

\section{Comparison with PIGS low-/medium-resolution}\label{sec:compAAT}

\begin{figure}[h]
\includegraphics[width=0.5\textwidth]{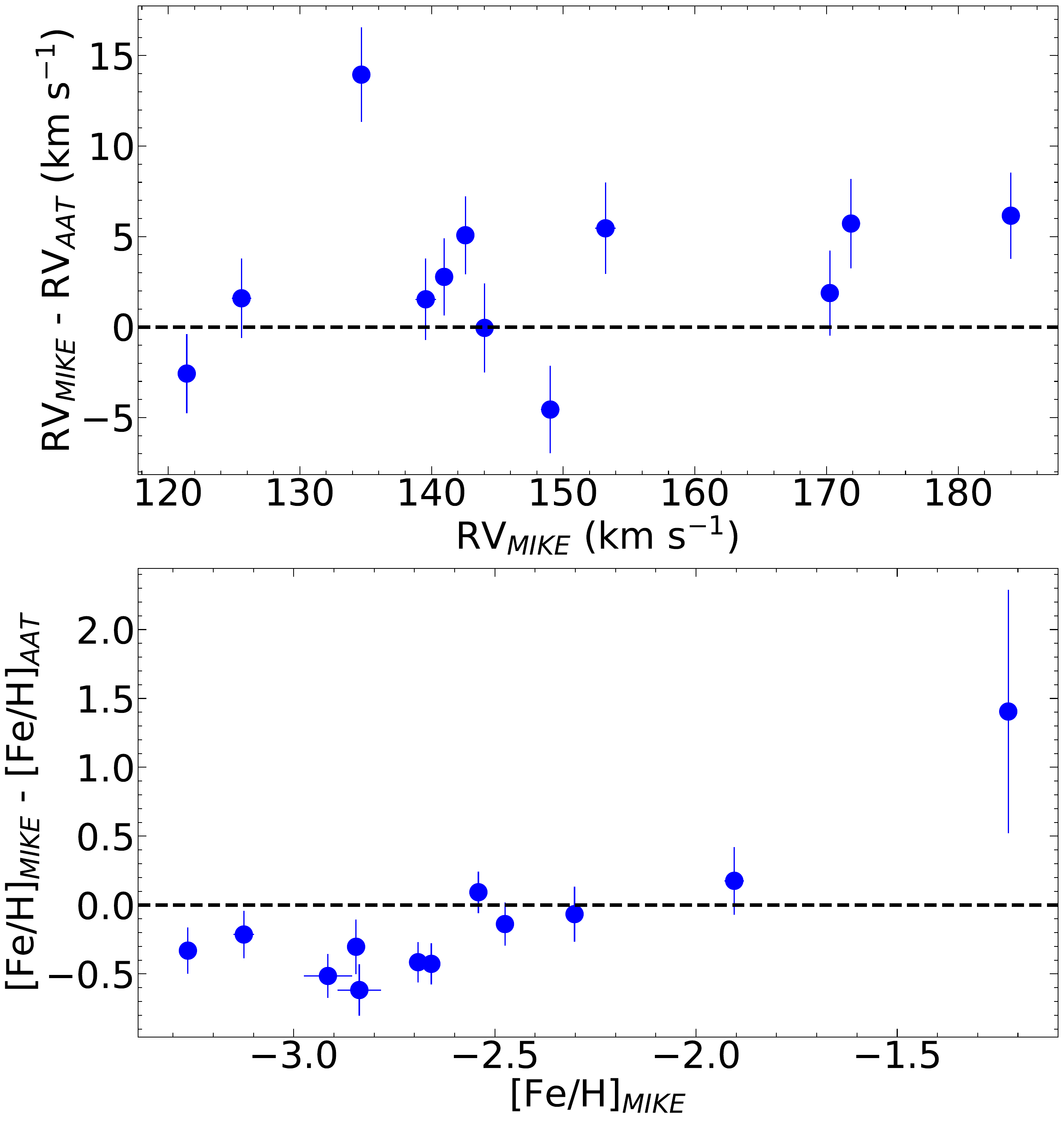}
\caption{Comparison of the radial velocity and \FeH{} for our targets between this work and PIGS/AAT. A systematic error of $2\kms$ has been added to the \texttt{FERRE} RV uncertainties as discussed in \citet{Arentsen24}.}
\label{Fig:comparison}
\end{figure}

The comparisons between the radial velocities and metallicities derived in this study versus the previous values obtained from the PIGS/AAT analysis \citep{Arentsen20b} are shown  in the top and bottom panels of Figure~\ref{Fig:comparison}, respectively. In the top panel, the radial velocities from both works show good agreement within $\sim5\kms$, with the exception of P185855$-$301522 that exhibits a difference of approximately $\sim15\kms$. The S/N measured in the AAT spectrum was similar to that of the other targets, $\sim62$ at 5000-5100\AA, hence this deviation suggests the possibility of P185855$-$301522 being part of a binary system, which is further discussed  in Section~\ref{sec:chem}. 

In the bottom panel, for the majority of stars in this sample, the difference in \FeH{} is below 0.5 dex, although the values obtained in this work tend to be systematically lower than those from the previous AAT analysis. This can be explained by different assumptions on the stellar parameters, by different models and techniques adopted to measure the \FeH{}, and by the different resolution of the instruments. 

Notably, one star, P185053$-$313317, exhibits a derived metallicity from the MIKE spectrum that is $\sim1.5$ dex higher than the value from the AAT. However, the uncertainty associated with the $\FeH_{\rm AAT}$ for this star is high ($0.88$ dex), weakening the discrepancy. This target represents the most metal-rich star in our sample ($\FeH_{\rm MIKE} \sim-1.2$). Given that the spectrum of this star provides measurements from up to $\sim175$ Fe\ione{} lines, the metallicity derived in this work is considered more reliable\footnote{We find that the PIGS photometric metallicity of the star is consistent with the low AAT metallicity. After carefully investigating, we conclude that the MIKE spectrum has to be of the same target and is not contaminated by any other sources, and our results are reliable. We will investigate this star in the future.}.

\section{Discussion}\label{sec:discussion}

\subsection{Distribution in metallicity and radial velocity}\label{sec:member}
\begin{figure}[h]
\centering
 \includegraphics[width=\columnwidth]{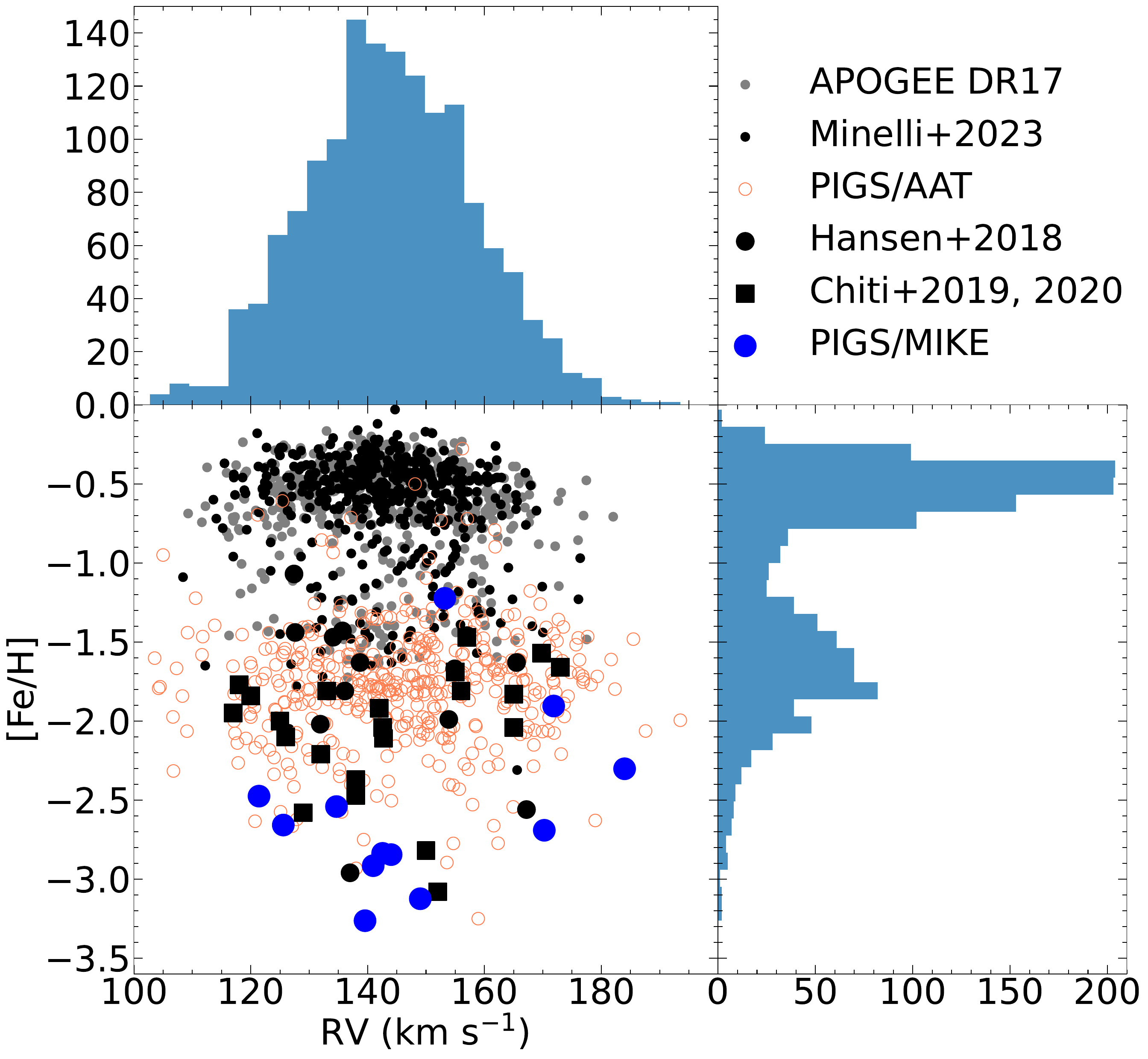} 
 \caption{Metallicities and radial velocities of Sgr members. Bottom left panel:  Radial velocities vs spectroscopic metallicities for Sgr stars from APOGEE DR17 (grey circles), from PIGS/AAT (open coral circles), from \citet[][small black circles]{Minelli23}, from \citet[][black circles]{Hansen18Sgr}, from \citet[][]{Chiti19} and \citet{Chiti20Sgr} with black squares, and from this work (blue circles). APOGEE DR17 stars have been selected imposing a S/N~$>70$ and to be spatially and kinematically in the core of the system. Top panel: Histogram of the radial velocities. Bottom right panel: Histogram of the metallicities.}
 \label{fig:histos}
\end{figure}

 The PIGS/MIKE radial velocities (RV) and  metallicites (blue circles) are displayed in Figure~\ref{fig:histos}, alongside with a similar selection in proper motion, radial velocity and position of Sgr members from APOGEE DR17 \citep[grey circles,][]{APOGEEDR17}, from \citet[][small black circles]{Minelli23}, from the PIGS/AAT (open coral circles) selected as in \citet{Vitali22} and \citet{Sestito24Sgr2}, and low-metallicity studies from \citet[][black circles]{Hansen18Sgr}, and with black squares from \citet{Chiti19} and  \citet{Chiti20Sgr}. 

The RV measured in our MIKE targets falls within the distribution of other Sgr stars, ranging between $100-200\kms$, with the mean occurring around $\sim 148\kms$ in our sample. This is compatible with  previous radial velocities studies of Sgr \citep[\eg][]{Ibata1994,Bellazzini2008,Minelli23,Sestito24Sgr2}.

The metallicity distribution of our targets  clearly stands out from previous analysis as it explores the low-metallicity region, where only $\sim5$ non-PIGS stars were studied so far with $\FeH\leq-2.5$ \citep[black markers,][]{Chiti20Sgr,Hansen18Sgr}. Therefore, the double peaked metallicity distribution in Figure~\ref{fig:histos} is caused by the various selection functions in the different datasets. In this work, we more than doubled the number of  stars (9 from MIKE, excluding PIGS/AAT members) with measured RV and \FeH{} from high-resolution spectroscopy  in this low-metallicity region. The metal-poor peak in the bottom-right panel  ($\FeH\sim-1.8$) clearly highlights the high efficiency of  PIGS in discovering metal-poor stars even towards the inner Galaxy, where the crowding of metal-rich stars  and the high extinction make Galactic archaeology studies harder than the normal halo.

\subsection{The very metal-poor tail of Sagittarius}\label{sec:chem}
\begin{figure*}
\includegraphics[width=\textwidth]{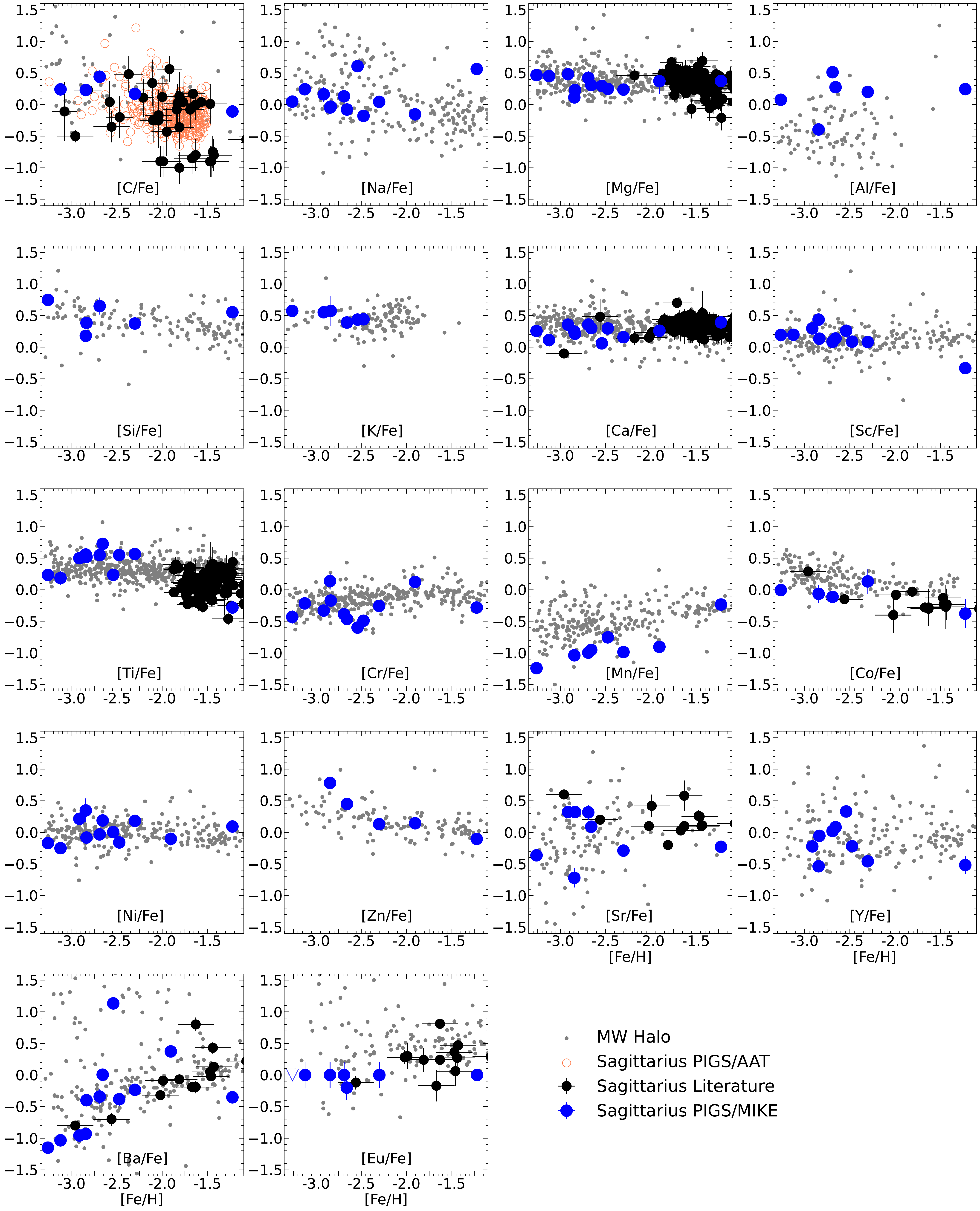}
\caption{Chemical abundances [X/Fe] as a function of [Fe/H]. Sgr stars analysed in this work are marked with blue circles. For [C/Fe], values are corrected for stellar evolution effects as in \citet{Placco14}. MW halo stars are from SAGA database \citep{Suda08} and they are denoted by  grey small circles. Literature stars of Sgr (black circles) for the various [X/Fe] are from \citet{Monaco05}, \citet{Mucciarelli17}, \citet{Hansen18Sgr}, except  for [C/Fe] which are from \citet{Hansen18Sgr}, \citet{Chiti19}, \citet{Chiti20Sgr}, and \citet{Sestito24Sgr2}. All datasets are in LTE. Halo stars have been selected to be all giants for a more fair comparison. Open triangle in our Eu measurement denotes an upper limit.}
\label{Fig:chems}
\end{figure*}

The chemical abundances of our targets are compared in Figure~\ref{Fig:chems} with a compilation of VMP stars from both the MW halo and Sgr. For the halo (grey circles), stars from the Stellar Abundances for Galactic Archaeology database\footnote{\url{http://sagadatabase.jp}} \citep[SAGA,][]{Suda08} are selected, including only stars with small uncertainties on [X/Fe] ($\lesssim 0.1$ dex).

Sgr stars (black markers) are selected from various spectroscopic studies, with observations targeting stars with $\FeH\lesssim-1.1$. As clearly shown in Figure~\ref{Fig:chems}, detailed chemical abundance analyses for VMP in Sgr are scarce and often limited to few atomic species.

The only study presenting a detailed chemical abundance analysis of VMP stars in this system and only for a restricted set of species (C, Ca, Co, Sr, Ba, La, Ce, Nd, Eu, Dy, Pb, and Th) is \citet{Hansen18Sgr}. Their sample comprises 13 stars (black circles), with only 4 being VMP ($\FeH\lesssim-2.0$) and just 2 falling within the range of $-3.0<\FeH<-2.5$. The authors suggest that a mixture of low- and high-mass Asymptotic Giant Branch (AGB) stars and supernovae is necessary to explain the observed chemical abundances \citep{Hansen18Sgr}. The measured level of [Ca/Fe] has been attributed by \cite{Hansen18Sgr} to the presence of  massive supernovae that enriched the early interstellar medium of Sgr before its gas was lost. 

Another two studies at intermediate-resolution, \cite{Chiti19} and \citet{Chiti20Sgr}, provide only \FeH{} and [C/Fe] for 4 and 18 stars (black squares), respectively. The former includes 2 stars in the range $-2.5<\FeH<-2.0$ and none below, while the latter consists of 9 VMPs, including one extremely metal-poor star (EMP, $\FeH<-3.0$). They found that none of their observed stars is Carbon-enhanced (CEMP), which contrasts with the fraction of CEMP stars observed in the Milky Way halo  \citep[][and references therein]{Arentsen22}. However, their results might be biased against CEMPs, since these stars can appear more metal-rich if selected with photometric metallicities  as discussed in \citet{DaCosta19} for the SkyMapper filter.

Data from the PIGS medium-resolution campaign (PIGS/AAT, open coral circles) are added as selected in \citet{Sestito24Sgr2}. This sample consists of 356 stars with $\FeH\leq-1.5$ and measured [C/Fe]. Among this sample, the PIGS/AAT [C/Fe] ratios indicate the presence of 3 CEMP stars, likely due to binarity. While PIGS photometry in Sgr is biased against CEMP stars, C-normal stars should not be affected by any selection biases. \citet{Sestito24Sgr2} discusses that the level of Carbon in Sgr is similar to the overall distribution of the other classical dwarf galaxies and lower than the level of the inner Galaxy and the MW halo. As also discussed by \citet{Vanni23}, this disparity is attributed to the higher efficiency of the interstellar medium (ISM) in DGs in retaining the ejected yields from the more energetic SNe events, which would produce more Fe than C, vs the lower efficiency in the smaller building blocks of the MW. This effect would also produce a lower CEMP-no fraction in DGs  \citep{Lucchesi24}, that is  CEMP linked to the supernovae from the First stars.

Only three stars from \citet{Monaco05} are included, as they have metallicities falling within the range of $-1.5<\FeH<-1.1$, while the remainder have higher metallicities. \citet{Mucciarelli17} analysed 235 Sgr stars, however only 124 have $\FeH\lesssim-1.1$, with  one being VMP ($\FeH\sim2.18$). These works (black circles) provide measurements only for [Mg, Ca, Ti/Fe]. The stars analysed in \citet{Mucciarelli17} are likely members of  the massive globular cluster M 54 and  the central nucleus of the galaxy (Sgr, N).

The data from Sgr exhibit good agreement in the [X/Fe] ratios,  with  stars from the literature being more metal-rich than our sample. Notably, the [C/Fe] ratios from \citet{Hansen18Sgr} are markedly lower than those from other datasets. This discrepancy is unlikely due to different populations in Sgr or stochasticity in the mixture of Type II supernovae (SNe~II); rather, it is more likely attributable to different assumptions on the stellar atmosphere models, atomic data, lines list, and the $\rm{^{12}C/^{13}C}$ isotopic ratio.

The general trend of the [X/Fe] ratios is similar between Sgr and MW halo stars for most elements. However,  [Al/Fe] appears to be higher in Sgr than in MW halo stars at the same metallicity (both datasets in LTE), with the latter generally exhibiting sub-solar values in the VMP region. Al is dispersed in the ISM by SNe~II and its yield is metallicity-dependent \citep{Woosley95,Nomoto13}. Additionally, the quantity of Al production is strongly influenced by the initial content of C and N in the gas cloud that formed the stars \citep{Kobayashi06}, hence, Al should correlate with C$+$N \citep[see also][]{Hawkins15,Das20}. Unfortunately, N is not measurable in our spectra, whilst [C/Fe] is measured only in 3 VMPs for which [Al/Fe] is also available, making it difficult to  establish any correlation in this sample. Another potential contributor to Al production  is from AGBs \citep{Nomoto13}, a topic discussed further in Section~\ref{sec:compdwarf}.

Conversely, [Mn/Fe] and [Co/Fe] in Sgr are  lower than the values observed in the MW halo.  The Sgr's deficiency in Mn and Co has been previously noted in various works at higher metallicities  \citep{McWilliam03,McWilliam13,Hasselquist17}. The level of Mn reflects the different star formation histories among different systems, and its deficiency has also been found in various classical DGs \citep[\eg][]{North12}. The lower Mn and Co abundance in Sgr compared to the MW halo has been attributed to the presence of yields from mass-dependent SNe~II \citep{Hasselquist17}. The lower  mass of SNe~II results in a lower  [Mn, Co/Fe]  in the subsequent generation of stars \citep{Chieffi04}. Additionally, \citet{Hasselquist17} pointed out that theoretical yields cannot fully reproduce the [Co/Fe] pattern,  however, Co can also be produced with a lower ratio than Fe in SNe~Ia, leading to a decrease in the [Co/Fe] ratio. 

The presence of SNe~Ia would produce a net decrease in the distribution of the alpha-elements' [X/Fe] ratios, resulting in the so-called $\alpha-$knee. The absence of an $\alpha-$knee in Sgr is evident for $\FeH<-2.0$, while it is challenging to exclude it at higher metallicities due to data limitations. However, the relatively low [Co/Fe] ratio at $\FeH>-2.0$ might suggest a possible contribution from SNe~Ia. This would imply a contribution of SNe~Ia at a lower metallicity than what previously found, $\FeH\sim-1.0$  \citep{Hasselquist17}.  The metallicity at which such a feature appears provides insights into the system’s efficiency in retaining metals in its interstellar medium \citep[\eg][]{Matteucci03,Venn04,Tolstoy09}. Other classical dwarfs exhibit an $\alpha-$knee in the range $-2.1<\FeH <-1.5$, depending on the  $\alpha$  used to trace the knee \citep{Reichert20,Sestito23Umi}. Recently, \citet{Sestito23Umi} precisely measured the  $\alpha-$knee for Ursa Minor (UMi) using APOGEE DR17 data, showing in that case that the knee occurs at $\FeH\sim-2.1$. The absence of the knee in Sgr at VMP metallicities suggests slightly higher enrichment efficiencies than UMi, while we cannot draw a more firm comparison from this work with the other DGs. This is because of the lack of data between $-2.0\lesssim\FeH\lesssim-1.0$. This chemical feature will be further investigated in Vitali et al. (in prep.).

The Ba panel in Figure~\ref{Fig:chems} shows that there is one Ba-rich star ([Ba/Fe]~$\sim1.2$), P185855$-$301522.  According to the PIGS/AAT medium-resolution observations \citep[][]{Arentsen20b}, this star qualifies as a Carbon-Enhanced Metal-Poor (CEMP, $\rm{[C/Fe]}=+0.97\pm0.24$, corrected for evolutionary effects). Unfortunately, C is not measurable in the MIKE spectrum of this star due to noise in the inter-order spectral region. Given its Ba- and C-rich nature, this star can be classified as CEMP-s \citep{Beers05}, and it is likely in a binary system. In support of this interpretation, the RV measurement from MIKE differs by $\sim15\kms$ from the AAT measurement.

\subsection{Comparison with other dwarfs and the presence of both r- and s-processes}\label{sec:compdwarf}

\begin{figure*}[h!]
\includegraphics[width=1\textwidth]{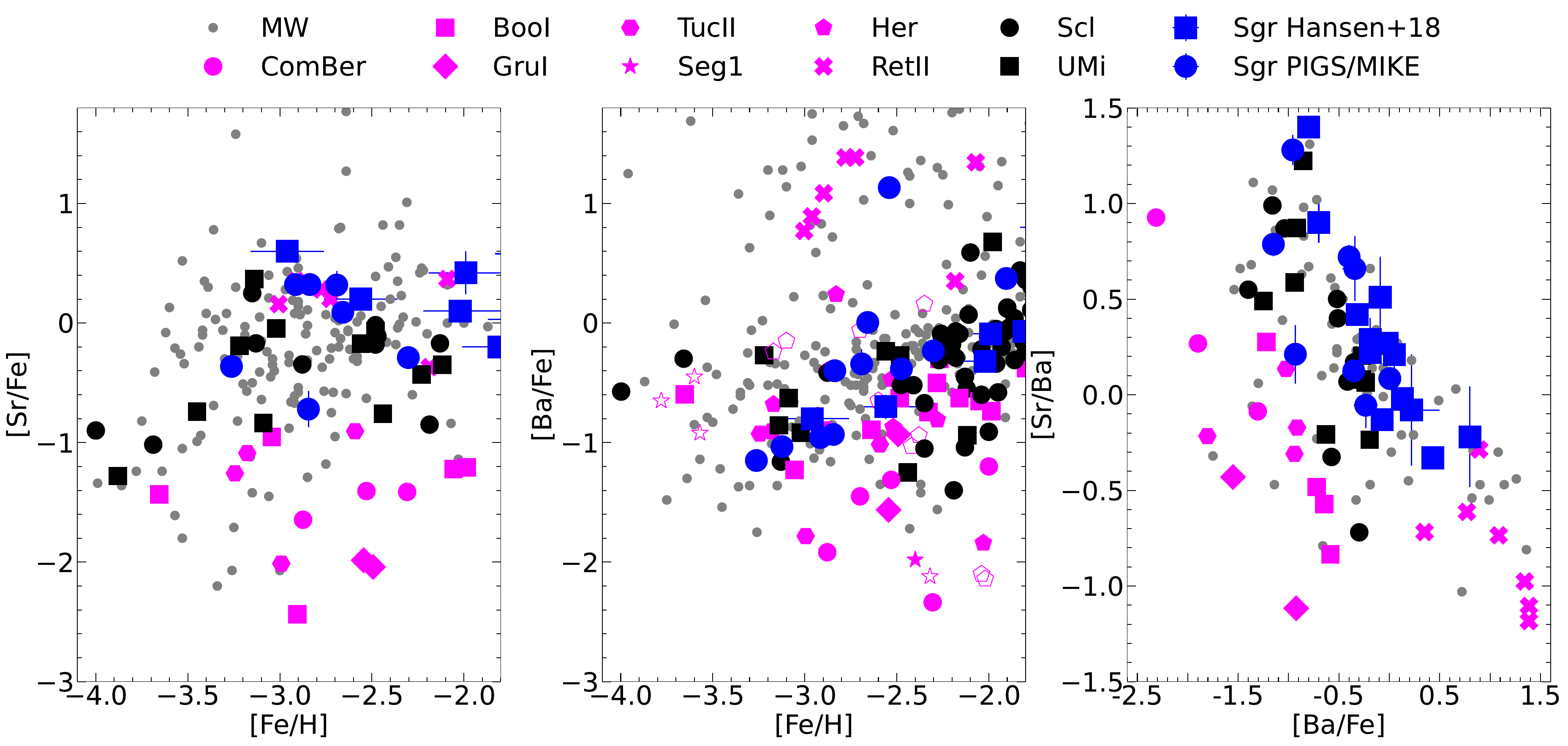}
\caption{Comparison of the neutron-capture elements. Left panel: [Sr/Fe] vs \FeH{}. Central panel: [Ba/Fe] vs \FeH{}. Empty symbols denote upper limits on the vertical axis. Right panel: [Sr/Ba] vs [Ba/Fe]. Grey circles are MW stars from  SAGA  \citep{Suda08}. Magenta markers denote UFDs: Coma Berenice (ComBer) stars are from \citet{Frebel10b} and \citet{Waller23}; Bootes~I (BooI) stars are from \citet{Feltzing09}, \citet{Norris10}, \citet{Gilmore13}, \citet{Ishigaki14}, and \citet{Frebel16}; Gru~I  stars are from \citet{Ji19}; Tucana~II (TucII) stars are from \citet{Ji16} and \citet{Chiti18}; Segue~1 (Seg1)  stars are from \citet{Frebel14}; Hercules (Her) stars are from \citet{Koch08}, \citet{Koch13}, and \citet{Francois16}; Reticulum~II (RetII) are from \citet{Ji16b} and \citet{Roederer16}. Black markers indicate classical DGs: Sculptor (Scl) are from \citet{Mashonkina17b} and \citet{Hill19}, while Ursa Minor (UMi) stars are from \citet{Mashonkina17b} and \citet{Sestito23Umi}, both DGs have been complemented with data from SAGA. All the datasets are from 1D LTE analyses. Uncertainties for stars from the literature are on the same scale as our values.}
\label{Fig:srba}
\end{figure*}

\begin{figure}[h!]
\includegraphics[width=0.5\textwidth]{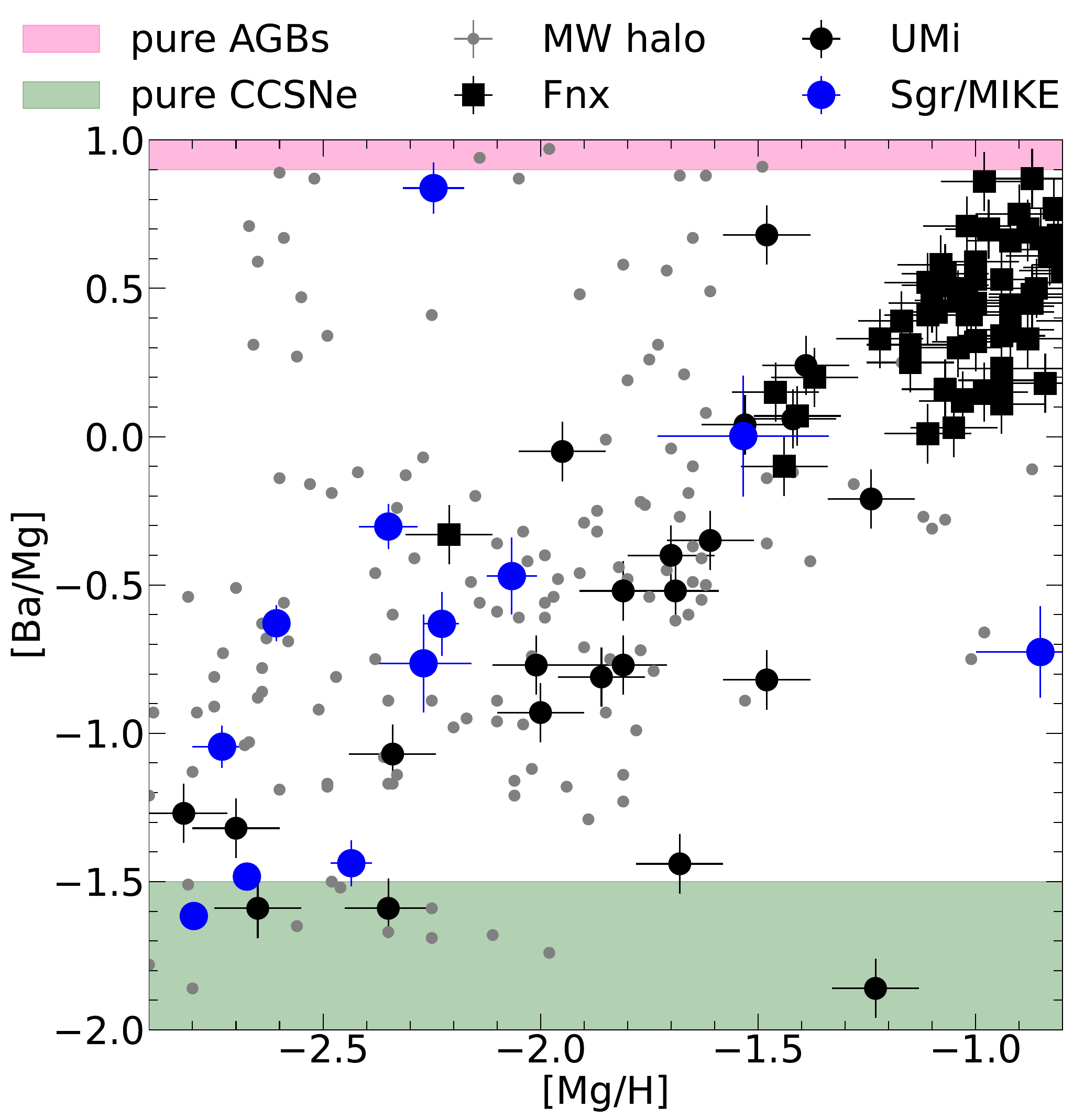}
\caption{[Ba/Mg] vs [Mg/H]. UMi stars (black circles) are from \citet{Shetrone01}, \citet{Sadakane04}, \citet{Cohen10}, \citet{Kirby12}, \citet{Ural15}, and \citet{Sestito23Umi}. Fnx stars (black squares) are from \citet{Letarte10}. MW halo stars (grey circles) are from the SAGA database. The green shaded area delimits the region where [Ba/Mg] is purely produced by core-collapse SNe~II (CCSNe), according to low-metallicity CCSNe yields from \citep{Ebinger20}. The pink shaded area indicates the region where [Ba/Mg] is produced by AGBs only, according to the theoretical yields from the \textsc{F.R.U.I.T.Y.} database \citep{Straniero06,Cristallo07,Cristallo09}.}
\label{Fig:bamg}
\end{figure}

\begin{figure}[h!]
\includegraphics[width=0.5\textwidth]{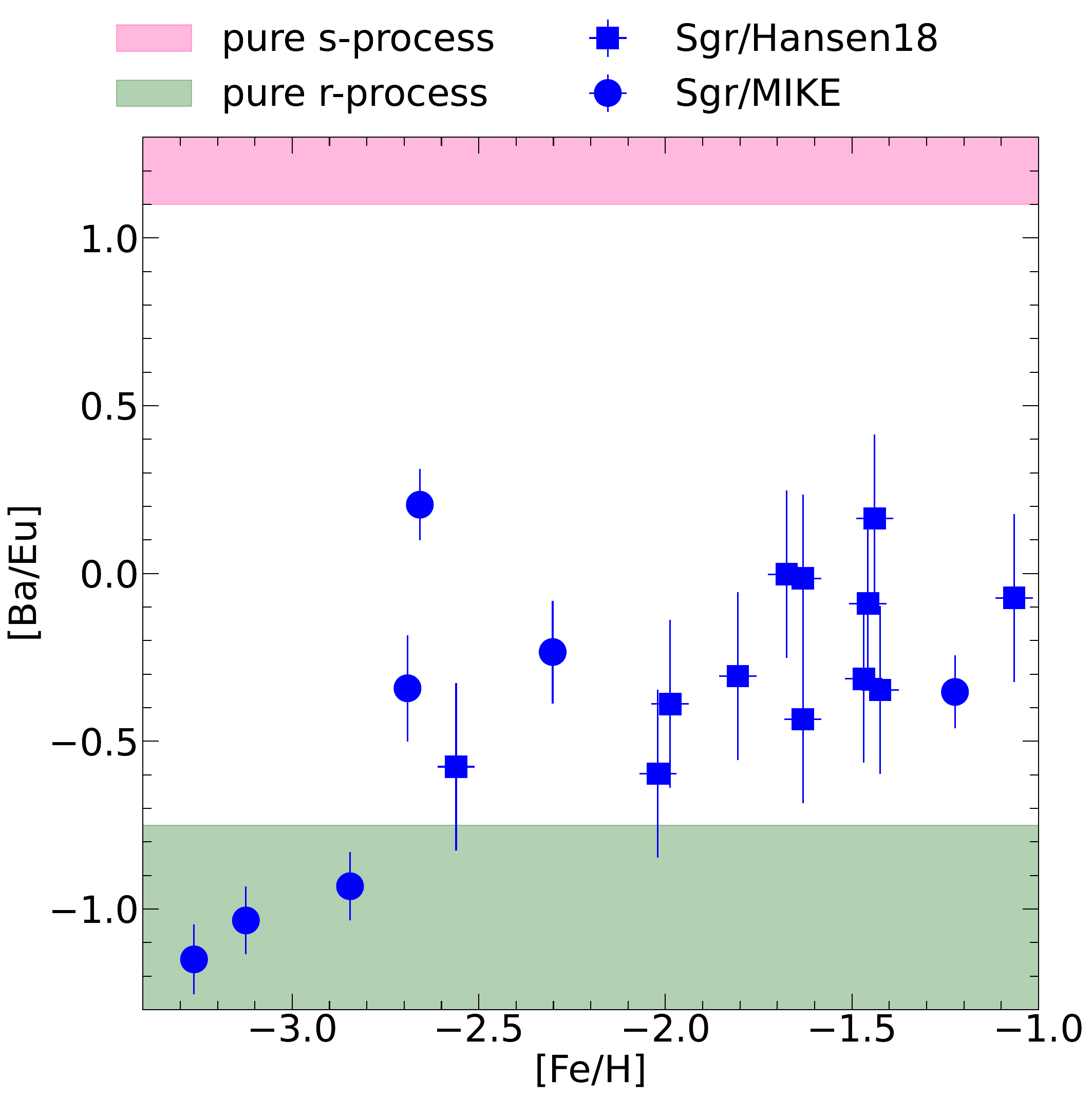}
\caption{[Ba/Eu] vs [Fe/H]. Blue circles and squares denote Sgr members from this work and from \citet{Hansen18Sgr}, respectively. The green shaded area delimits the region where [Ba/Eu] is purely produced by r-process channels, while the pink shaded area indicates the region where this ratio is produced by AGBs only \citep{Arlandini99,Bisterzo14}.}
\label{Fig:baeu}
\end{figure}

\begin{figure}[h!]
\includegraphics[width=0.5\textwidth]{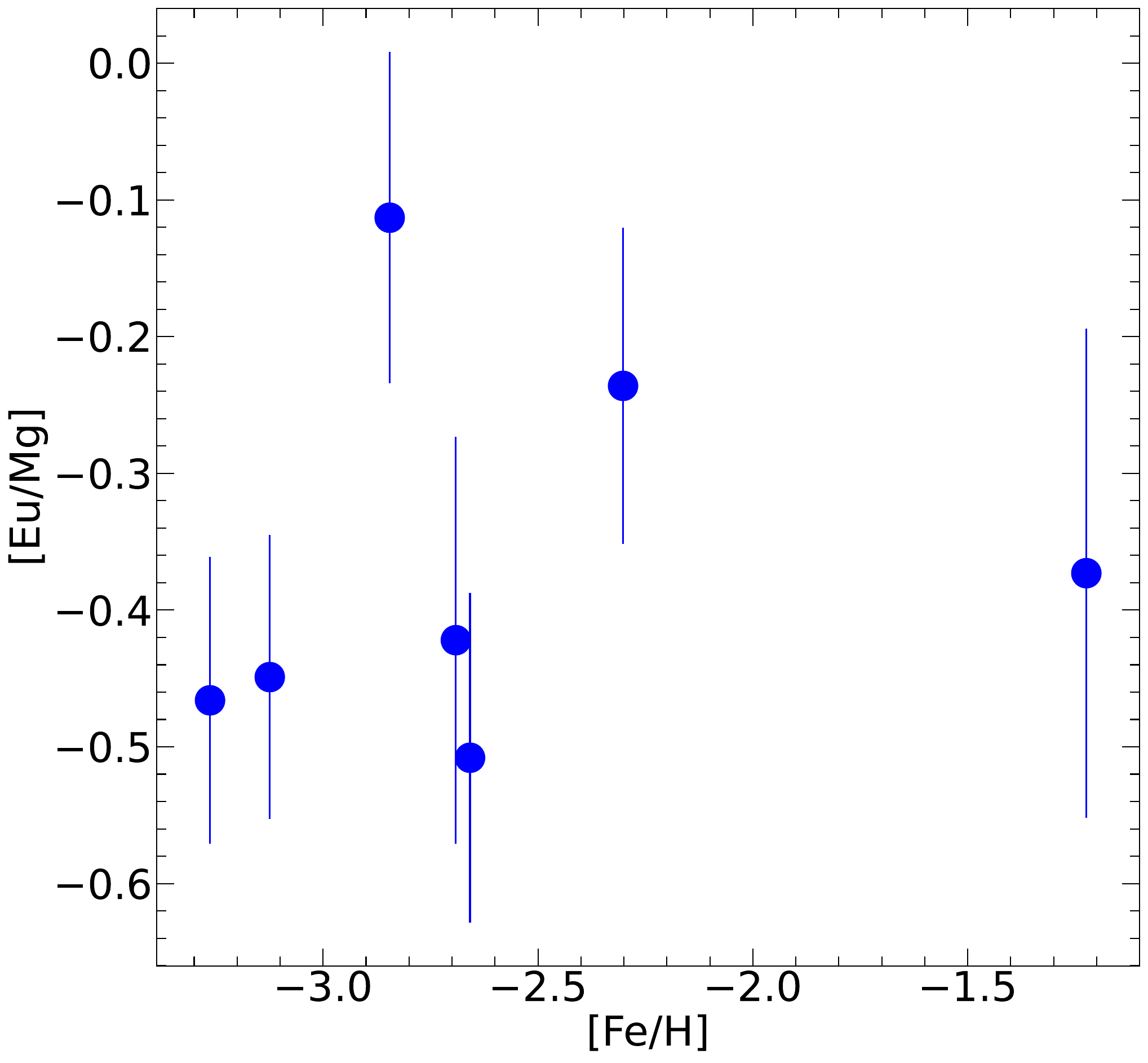}
\caption{[Eu/Mg] vs [Fe/H]. Blue circles denotes Sgr members from this work.}
\label{Fig:eumg}
\end{figure}

A comparison of the neutron-capture elements in Sgr with other dwarf (black markers) and ultra-faint dwarf (magenta markers) galaxies is shown in Figure~\ref{Fig:srba}. Our work and the Sgr stars analysed in \citet{Hansen18Sgr} are represented by blue markers. The complete list of works for the other systems used for comparison is provided in the caption of Figure~\ref{Fig:srba}. 

As expected, given the mass of the system and its extended star formation history, Sgr exhibits a similar chemical abundance ratio of [Sr, Ba/Fe] as the other classical DGs (left and central panels), which is higher than that observed in ultra-faint systems \citep[\eg][on the level of n-capture elements in UFDs and DGs]{Ji19}. This similarity suggests that multiple nucleosynthetic channels likely contributed to enriching the interstellar medium with these heavy elements. The wide spread in  [Sr/Ba] distribution (right panel) and the super-solar value for some Sgr stars indicate that both slow- and rapid-neutron capture processes played a role in chemical enrichment,  particularly in forming more  Sr than Ba \citep[\eg][]{Mashonkina17b}. 

One star, P185704$-$301021, is VMP and  has a high [Sr/Ba] ratio ($\sim 1.3$), similar to a star analysed by \citet{Hansen18Sgr}, Sgr~2300275. They interpreted the chemical signature of this star as  originating from  pure r-process, given its [Ba/Eu] ratio that may suggest this conclusion \citep{Hansen18Sgr}. However, their Eu measurement is only available as an upper limit, resulting in a lower limit for their [Ba/Eu], implying that the contribution of s-process channels cannot be excluded. In fact, s-processes from low-metallicity fast-rotating massive stars can also  explain an excess of Sr-production over Ba \citep{Pignatari08}.

The chemical space [Ba/Mg] vs [Mg/H] is able to discern if a contribution from  AGB metal-poor stars is present in a closed system other than SNe~II. This chemical space is displayed in Figure~\ref{Fig:bamg}, showing our sample, stars from Ursa Minor (UMi), from Fornax (Fnx), and from the MW halo. A scenario where only SNe~II contribute, that is  primarily r-process, would produce a flat distribution of [Ba/Mg], also known as a Ba-floor \citep[\eg][and references therein]{Cowan21,Mashonkina22}. This Ba-floor is shown with the green shaded area, which is obtained from theoretical low-metallicity supernovae yields models \citep{Ebinger20}. On the other hand, if AGBs are present, they would produce more Ba than Mg through s-processes \citep[\eg][]{Pignatari08,Cescutti14}, leading to an increase in [Ba/Mg] as [Mg/H] increases. This pure-AGB scenario is depicted by the pink shaded area, derived from AGB models from the  \textsc{F.R.U.I.T.Y.} database\footnote{\url{http://fruity.oa-teramo.inaf.it/modelli.pl}} \citep{Straniero06,Cristallo07,Cristallo09}.

In both UMi and Sgr, the bottom of the [Ba/Mg] distribution is approximately $\sim-1.5$, while it is around $\sim-0.3$ for Fnx. The higher minimum level of [Ba/Mg] in Fnx is because the sample is more metal-rich \citep{Letarte10} than the stars in Umi and Sgr. However, this does not influence the interpretation of this chemical space. As [Mg/H] increases, this ratio rises by up to $\sim2.5$ dex in Sgr and Umi and by up to $\sim1.2$ dex in Fnx, a significant deviation beyond a $5\sigma$ error. Recently, the spread in the [Ba/Mg] in UMi has been reported by \citet{Sestito23Umi}, concluding that contribution of metal-poor AGBs in UMi is needed to explain such chemical trend. In the case of Fnx, \citet{Letarte10} found that the s-process channels in stellar winds from low-metallicity AGBs are needed to reproduce the chemical pattern of stars in that system. Similarly, we can conclude that AGBs are required in the chemical evolution of Sgr, as also suggested by  \citet{Hansen18Sgr}. Contribution from AGBs in DGs has been found to be common and they are expected to have had a greater relative contribution to the systems enrichment than in the MW halo \citep[][]{Hasselquist21}. 

Another similar diagnostic for s-process channels is to use the [Ba/Eu] ratio \citep[\eg][]{Mashonkina17b}. Eu is purely produced by r-processes, therefore an increase of [Ba/Eu] as a function of [Mg or Fe/H] would indicate a contribution from s-channels to account for the extra Ba. The [Ba/Eu] vs \FeH{} space is displayed in Figure~\ref{Fig:baeu}, reporting the 6 MIKE stars with Eu measurement and 12 Sgr stars from \citet{Hansen18Sgr}.  Both datasets show a rise in the [Ba/Eu] of $\sim1.5$ dex, confirming the contribution of AGBs in the chemical evolution of Sgr.

Stars studied here are supposed to be old as expected for VMP populations. \citet{Siegel07} found Sgr stars with $\FeH\sim-1.2$ to be older than 10 Gyr. However, we note that the stellar ages should be limited to allow AGBs to enrich the ISM.

The r-process yields can be produced by either a prompt nucleosynthetic channel, such as SNe~II, or by a delayed events, that is  compact binary merger. In case of prompt event, the [Eu/Mg] would be mostly independent by metallicity, that is  Eu and Mg would be produced by the same nucleosynthetic channels. Otherwise, a rise of [Eu/Mg] with \FeH{} would indicate an extra channel for Eu occurred at a later time \citep[\eg][]{Skuladottir20}. This chemical space, [Eu/Mg] vs \FeH{},  is shown in Figure~\ref{Fig:eumg}, displaying only stars from our MIKE analysis, which is the only high-resolution analysis providing both Eu and Mg for stars in the Sgr's core at these metallicities. Our measurements of [Eu/Mg] shows that there is no clear trend with \FeH{}, however, the sample in Figure~\ref{Fig:eumg} is relatively small and it will be extended at higher metallicities by Vitali et al. (in prep.). The large dispersion might suggest a partial contribution from  delayed r-process sources,  such as compact binary mergers. In the Section below, we discuss the possible SNe~II contributions using a best fit tool for theoretical yields, including compact binary events.

\subsection{Type II supernovae enrichment in Sgr}\label{sec:yields}
Sgr is an evolved and massive system similar to other classical dwarfs, implying that multiple SNe~II contributed to its chemical enrichment history. To gain insight into the specific types of SNe~II needed to reproduce the chemical properties of the VMP tail of Sgr, \textsc{StarFit}\footnote{\url{https://starfit.org}} is employed. This tool  provides a best fit, which is a combination of various SNe~II yields selected by the user from a pool of theoretical models. Ten models were chosen, encompassing a wide variety of supernova events, including faint SNe, core-collapse, high-energy, hypernovae, rotating massive stars, compact binary mergers, pair-instability SNe, and mixing both Population~III~and~II stars.

\begin{figure*}[h!]
\includegraphics[width=\textwidth]{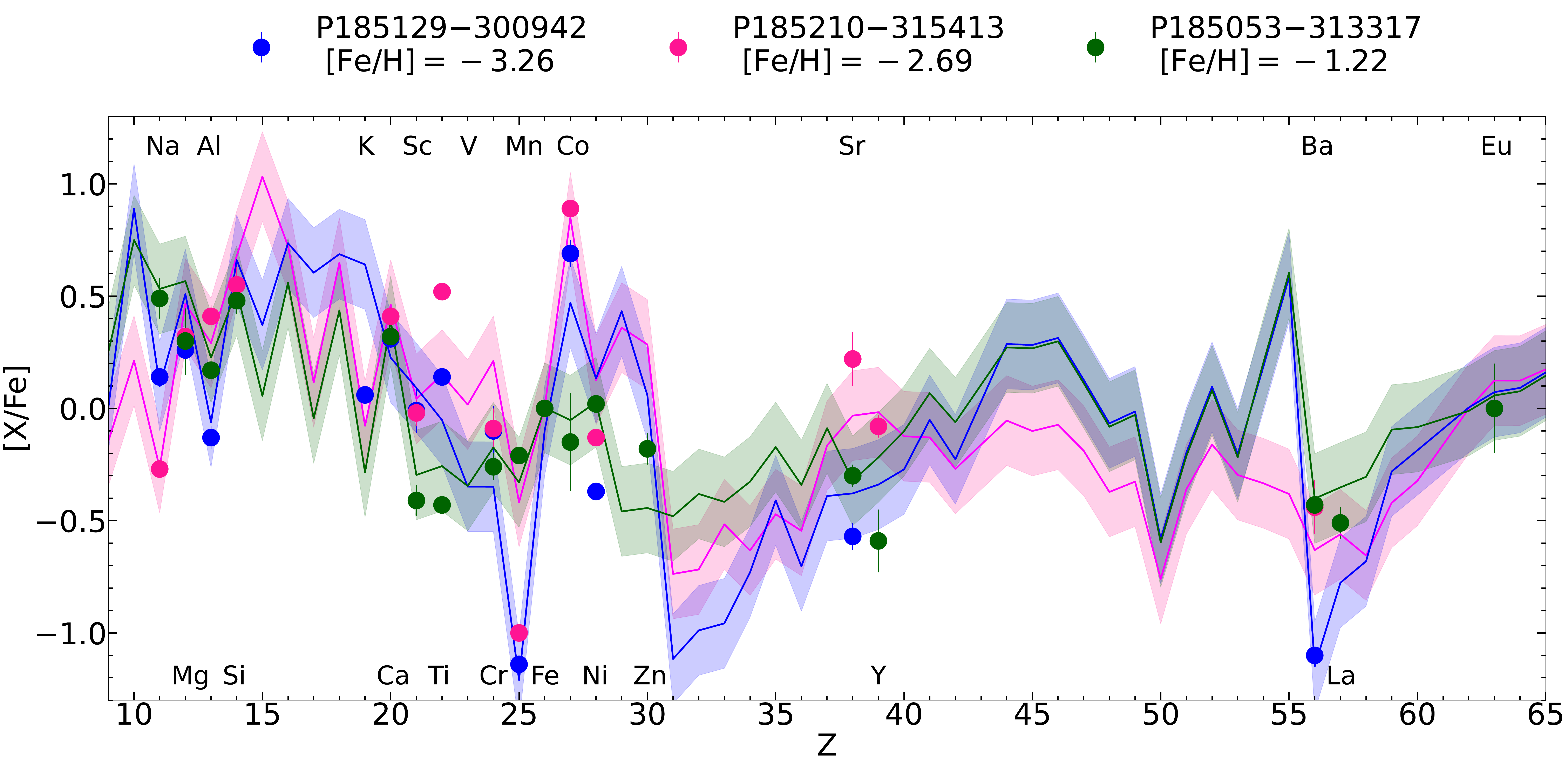}
\caption{Chemical abundance ratio [X/Fe] as a function of the proton number Z. Measured [X/Fe] for P185129$-$300942, P185210$-$315413, and P185053$-$313317  are reported with blue, deep pink, and red  circles, respectively. Theoretical [X/Fe] ratios from the \textsc{StarFit} supernovae yields best fits are shown with solid lines coloured as the observed data. Arbitrary uncertainties ($\pm0.2$) on the theoretical yields are represented with shadow areas. NLTE-corrected [X/Fe] are used for some species (see Section~\ref{sec:nlte}). The names of the atomic species detected in this work are reported in the plot.}
\label{Fig:starfit}
\end{figure*}

\textsc{StarFit} is run for 3 stars in our sample: P185129$-$300942 (the most metal-poor), P185210$-$315413 (a typical VMP), and P185053$-$313317 (the most metal-rich star in our MIKE analysis). The tool is used multiple times, imposing fits to the observed data with the mixing of 3, 5, and 8 supernova events. This exercise is repeated, first adopting LTE-only chemical ratios first and, then, using NLTE-corrected when available (see Section~\ref{sec:nlte}). 

It is important to note that our interpretation of the supernova events from \textsc{StarFit} is more qualitative, rather than quantitative, as there may be systematic differences in the models across various theoretical assumptions and complexities in the mixing of supernova ejecta into the interstellar medium of  massive systems as DGs. Additionally, it is worth mentioning that \textsc{StarFit} lacks yields from Type Ia supernovae and AGBs. The presence of SNe Ia will be investigated in a separate paper focusing on observations of more metal-rich stars (Vitali et al., in prep.).

The chemical composition of the stars in Sgr can be explained by a complex mixing of various SNe~II events. Both the LTE and the NLTE case provide the same qualitative picture. The best fits (solid lines) for the three stars provided by \textsc{StarFit} are displayed in Figure~\ref{Fig:starfit}, alongside  the chemical ratios of P185129$-$300942 (blue), P185210$-$315413 (deep pink), and P185053$-$313317 (red). Results are shown as [X/Fe] as a function of the proton number. 

To explain the level of the heavy elements, compact binary merger events are needed \citep{Just15}. This kind of delayed event is needed to likely account for the spread in the [Eu/Mg] ratio as shown in Figure~\ref{Fig:eumg}. Additionally, fast-rotating intermediate-mass and massive  metal-poor stars \citep[$\sim25-120\msun$, up to $\sim300\kms$,][]{Limongi18} must be included to explain the higher [Sr, Y/Fe] compared to [Ba/Fe]. The latter is an s-process channel that would also explain the high [Sr/Ba] stars found in this work and in \citet{Hansen18Sgr}. In concomitance with the output from \textsc{StarFit}, AGBs are needed to reproduce the [Ba/Mg] and [Ba/Eu] spread in Sgr (see Section~\ref{sec:compdwarf}).

To fit the lighter elements up to the Fe-peak, a mixture of low- to intermediate-mass high-energy SNe and hypernovae \citep[$\sim10-70\msun$,][]{Heger10,Grimmett18} are needed.

\section{Conclusions}\label{sec:conclusion}
The very metal-poor region of the Sagittarius dwarf galaxy has so far been poorly investigated. In this work, we provide the first and most detailed chemical abundance analysis for 12 stars, shedding light on the chemical evolution of this system. The targets were previously observed with AAT medium-resolution within the Pristine Inner Galaxy survey \citep{Arentsen20b}. Precise radial velocities and chemical abundances of up to 17 species are measured from the high-resolution spectra of the MIKE spectrograph at the {\it Magellan-Clay} telescope. We find that:

\begin{enumerate}
    \item RVs and metallicities from this work are in good agreement with the previous AAT analysis (see Figure~\ref{Fig:comparison}). 
    \item Only the RV of one star, P185855$-$301522, largely differs by $\sim15\kms$ from the previous measurement. Given its [C, Ba/Fe] ratios, this star is classified as  CEMP-s and likely in a binary system.
    \item The number of  stars in the core of Sgr with $\FeH\leq-2.5$ is more than doubled. So far only 5-6 stars were analysed providing RVs, metallicities and carbonicities \citep{Hansen18Sgr,Chiti18,Chiti20Sgr}, if PIGS/AAT are excluded. Out of the 12 stars here analysed, 9 have $\FeH\leq-2.5$ (see Figure~\ref{fig:histos}). 
    \item The chemical abundances of Sgr do not differ from the MW halo distribution for the majority of the species analysed in this work (see Figure~\ref{Fig:chems}). The exceptions are Mn and Co, which are at the low-end of the MW trend, while the level of Al is higher than the one in the MW. These differences can be explained by the mass- and metallicity-dependent SNe~II yields, and they are common in classical DGs.
    \item The abundances derived in this work are compatible with the trend of Sgr stars from the literature at higher metallicities (see Figure~\ref{Fig:chems}).
    \item There is no detection of an $\alpha-$knee in the VMP region, which implies a lack of SNe~Ia. However, the low [Co/Fe] at $\FeH>-2.0$ in comparison with the MW halo trend might also suggest a contribution from SNe~Ia. A coming PIGS paper (Vitali et al., in prep.) will investigate higher metallicities to better characterise the presence of type Ia supernovae in Sgr.
    \item The distribution of [Sr, Ba/Fe] and of the [Sr/Ba] in Sgr is similar as in other massive classical DGs (see Figure~\ref{Fig:srba}). The wide spread in [Sr/Ba] is indicative that both rapid- and slow-process channels are present in the chemical history of the system.
    \item The wide spread in [Ba/Mg] as a function of [Mg/H] and in  [Ba/Eu] vs[Fe/H] are indicative that asymptotic giant branch stars polluted the interstellar medium of the system (see Figures~\ref{Fig:bamg}~and~\ref{Fig:baeu}), as in other classical DGs.
    \item The chemical trend of these Sgr stars can be fitted by the mixture of multiple type II supernovae (see Figure~\ref{Fig:starfit}). In particular, to reproduce the lighter elements up to the Fe-peak, low- to intermediate-mass high-energy SNe and hypernovae ($\sim10-70\msun$) are needed. The trend of the heavy elements is explained by the presence of compact binary merger events (see also Figures~\ref{Fig:eumg}) and fast-rotating (up to $\sim300\kms$) intermediate-mass to massive metal-poor stars ($\sim25-120\msun$),  sources of rapid- and slow-processes, respectively.
\end{enumerate}

Altogether, this work provides an unprecedented detailed view of the very metal-poor population in Sgr, and early chemical enrichment of dwarf galaxies in general. A coming PIGS paper will investigate the more metal-rich population of Sgr (Vitali et al., in prep.), exploring the contribution from type Ia supernovae, while future spectroscopic surveys, for example 4DWARFS \citep{Skuladottir23}, will also be dedicated to dissecting the chemical properties of Sgr, its core and its stellar streams. It will be interesting to investigate the various populations inside this system and try to understand its formation and chemo-dynamical evolution.

\begin{acknowledgements}
We acknowledge and respect the l\textschwa\textvbaraccent {k}$^{\rm w}$\textschwa\ng{}\textschwa n peoples on whose traditional territory the University of Victoria stands and the Songhees, Esquimalt and WS\'ANE\'C  peoples whose historical relationships with the land continue to this day.
\\
\\
FS and KAV thank the National Sciences and Engineering Research Council of Canada for funding through the Discovery Grants and CREATE
programs. FS acknowledges the Millennium Nucleus ERIS (ERIS NCN2021017) for financially supporting his visit at the Universidad Diego Portales. SV thanks ANID (Beca Doctorado Nacional, folio 21220489) and Universidad Diego Portales for the financial support provided. SV, PJ, DdBS, and CJLE acknowledge the Millennium Nucleus ERIS (ERIS NCN2021017) and FONDECYT (Regular number 1231057) for the funding. FG gratefully acknowledges support from the French National Research Agency (ANR)- funded projects ''MWDisc'' (ANR-20-CE31-0004) and ''Pristine'' (ANR-18-CE31-0017). NFM gratefully acknowledges support from the French National Research Agency (ANR) funded project ''Pristine'' (ANR-18-CE31-0017) along with funding from the European Research Council (ERC) under the European Unions Horizon 2020 research and innovation programme (grant agreement No. 834148). ES acknowledges funding through VIDI grant "Pushing Galactic Archaeology to its limits" (with project number VI.Vidi.193.093) which is funded by the Dutch Research Council (NWO). This research has been partially funded from a Spinoza award by NWO (SPI 78-411). This research was supported by the International Space Science Institute (ISSI) in Bern, through ISSI International Team project 540 (The Early Milky Way).
\\
\\
This work has made use of data from the European Space Agency (ESA) mission {\it Gaia} (\url{https://www.cosmos.esa.int/gaia}), processed by the {\it Gaia} Data Processing and Analysis Consortium (DPAC, \url{https://www.cosmos.esa.int/web/gaia/dpac/consortium}). Funding for the DPAC has been provided by national institutions, in particular the institutions participating in the {\it Gaia} Multilateral Agreement.
\\
\\
This research has made use of the SIMBAD database, operated
at CDS, Strasbourg, France \citep{Wenger00}. This work made
extensive use of TOPCAT \citep{Taylor05}.
\end{acknowledgements}

\bibliographystyle{aa}
\bibliography{VMP_Sgr}

\end{document}